\documentclass[aps,prl,twocolumn,groupedaddress,showpacs,floatfix,superscriptaddress]{revtex4-1}
\usepackage{epsfig}
\usepackage{amsmath,amssymb}
\usepackage{graphicx}
\usepackage[dvipsnames,usenames]{color}
\usepackage[normalem]{ulem}
\usepackage{soul} 
\emergencystretch=\maxdimen
\begin{document}

\title{Competition between Phase Separation and Spin Density Wave or Charge Density Wave Order: Role of Long-Range Interactions}
\author{Bo Xiao}
\email {boxiao@ucdavis.edu} 
\affiliation{Department of Physics, University of California, 
Davis, CA 95616,USA}
\author{F. H\'ebert}
\affiliation{Universit\'e C\^ote d'Azur, CNRS, INPHYNI, France}
\author{G. Batrouni}
\affiliation{Universit\'e C\^ote d'Azur, CNRS, INPHYNI, France}
\affiliation{MajuLab, CNRS-UCA-SU-NUS-NTU International Joint Research
  Unit, 117542 Singapore}
\affiliation{Centre for Quantum Technologies, National University of Singapore, 2 Science Drive 3, 117542 Singapore} 
\affiliation{Department of Physics, National University of Singapore, 2
  Science Drive 3, 117542 Singapore}
\affiliation{Beijing Computational Science Research Center, Beijing
  100193, China}
\author{R.T. Scalettar}
\affiliation{Department of Physics, University of California, 
Davis, CA 95616,USA}

\begin{abstract}
Recent studies of pairing and charge order in materials such as FeSe,
SrTiO$_3$, and 2H-NbSe$_2$ have suggested that momentum dependence of
the electron-phonon coupling plays an important role in their
properties.  Initial attempts to study Hamiltonians which either do
not include or else truncate the range of Coulomb repulsion have noted
that the resulting spatial non-locality of the electron-phonon
interaction leads to a dominant tendency to phase separation.  Here we
present Quantum Monte Carlo results for such models in which we
incorporate both on-site and intersite electron-electron
interactions.  We show that these can stabilize phases in which the
density is homogeneous and determine the associated phase boundaries.
As a consequence, the physics of momentum dependent electron-phonon
coupling can be determined outside of the trivial phase separated
regime.
\end{abstract}

\date{\today}

\pacs{
71.10.Fd, 
71.30.+h, 
71.45.Lr, 
74.20.-z, 
02.70.Uu  
}
\maketitle

\section{1.  Introduction}

The challenging nature of the computational solution of the quantum
many electron problem has led, to a quite considerable extent, to
consideration of models which incorporate only a single type of
interaction.  For example, the Hubbard and periodic Anderson
Hamiltonians focus on on-site electron-electron repulsion, the Kondo
model on an interaction with local spin degrees of freedom, while the
Holstein and Su-Schrieffer-Heeger Hamiltonians consider exclusively
electron-phonon interactions.  This is, of course, an unfortunate
situation, since the interplay between different interactions can lead
to transitions between associated ordered phases which are of great
interest.  Even more importantly, this competition is present in most
real materials.  Quantum Monte Carlo (QMC) simulations of the
Hubbard-Holstein Hamiltonian in two
dimensions\cite{berger95,assaad07,nowadnick12,johnston13,yamazaki14,macridin12,mendl17,ghosh18,karakuzu17}
have reinforced the challenges of, as well as the interest in,
including multiple interactions.  A very recent study offers great
promise in developing a QMC approach in which electron-electron and
electron-phonon interactions can be folded together, at least in the
particular region of the phase diagram where charge order
dominates\cite{karakuzu18}.

One dimensional systems have offered an exception to this rule,
largely because specialized algorithms exist in cases when the
restricted geometry forbids the exchange of electrons\cite{hirsch82}.
Thus the extended Hubbard model (EHH), which includes
electron-electron interactions in the form of both an on-site $U$ and
an intersite $V$ is now known to include not only the expected charge
density wave (CDW) correlations for $U<2V$ and spin density wave (SDW)
correlations (for $U>2V$), but also a subtle bond ordered wave (BOW)
phase in a narrow region about the line $U=2V$
\cite{nakamura99,nakamura00,sengupta02,sandvik04,tsuchiizu02,tsuchiizu04,zhang04,tam06}.
These simulations have been used to understand better the behavior of
different materials, including organic
superconductors\cite{berlinsky79}, density wave
materials\cite{gruner94}, and undoped conducting
polymers\cite{bourbonnais08}.  Even with the applicability of these
one-dimensional approaches, however, the physics remains subtle, and
even controversial, owing mainly to energy gaps vanishing
exponentially in the limit of weak coupling.

The goal of the present paper is an investigation of the physics of a
Hamiltonian which includes long range electron-phonon coupling as well
as electron-electron interactions $U$ and $V$.  Motivation is given by
several recent situations in which momentum-dependent electron-phonon
coupling (which necessitates non-local coupling in real space) has been
suggested to be important, as we discuss below.  Our work builds on the
considerable earlier literature of one dimensional models with mixed
interactions, which we review.  
As a first step, we focus here on half-filling and 
the effects of coexistence of long range electron-phonon 
coupling and electron-electron interactions.  However, 
our numerical method, introduced in the following section, can be applied to 
this model at any fillings without a sign problem.

A further reason to focus on longer range interactions is in the search
for one dimensional models which exhibit superconductivity as their
dominant correlation.  Despite an initial suggestion to the
contrary\cite{clay05}, it now seems most likely that the one dimensional
Hubbard-Holstein model with only on-site electron-phonon coupling, does
not have dominant pairing fluctuations\cite{hardikar07,tam07}.  Instead,
an `extended Holstein' model, in which phonons residing on the midpoints
between lattice sites couple to the electronic charge both to the left
and to the right, has been put forth as a `minimal model' in which the
electron-phonon can dominate over the more typical gapped CDW and SDW
phases at half-filling\cite{bonca01,tam14}.  

We begin by writing down the full Hamiltonian,
and then consider the various limiting cases before
investigating the physics when all terms are present.
Our model is,
\begin{align} 
\hat {\cal H} &=
\hat {\cal K} + 
\hat {\cal H}_{ph}  + 
\hat {\cal P}_u +
\hat {\cal P}_v +
\hat {\cal P}_{ep}
\nonumber
\\
\hat {\cal K} &= 
-t \sum_{l, \sigma} 
\big( \, \hat c^{\dagger}_{l, \sigma} 
\hat c^{\phantom{\dagger}}_{l+1,\,  \sigma} + 
 \hat c^{\dagger}_{l+1, \, \sigma} 
\hat c^{\phantom{\dagger}}_{l, \sigma}  \big)
\nonumber \\
\hat {\cal H}_{ph} &= 
\frac{1}{2} \sum_{l}
\,\big(\, \omega^2 \hat X_{l}^2
+  \hat P_{l}^2 \,\big)
\nonumber \\
\hat {\cal P}_u &= U \sum_{l}
\hat n_{l, \uparrow} \, \hat n_{l, \downarrow} 
\nonumber \\
\hat {\cal P}_v &=  V \sum_{l}
\big( \, \hat n_{l, \uparrow} + \hat n_{l, \downarrow}  \, \big)
\big( \, \hat n_{l+1, \, \uparrow} + \hat n_{l+1, \, \downarrow}  \, \big)
\nonumber 
\\
\hat {\cal P}_{ep} &=  \sum_{l, r}
\lambda(r) \hat X_l 
\big( \, \hat n_{l+r, \, \uparrow} + \hat n_{l+r, \, \downarrow}  \, \big)
\label{eq:H}
\end{align}
Here $\hat {\cal H}$ is comprised of a kinetic energy ($\hat {\cal
  K}$) describing the hopping of fermions along a one-dimensional
chain of $N$ sites; an on-site repulsion $U$ between fermions of
opposite spin $\sigma$ ($\hat {\cal P}_u$); an intersite repulsion $V$
between fermions on adjacent sites ($\hat {\cal P}_v$); and, finally,
an electron-phonon coupling ($\hat {\cal P}_{ep})$.  All energies will
be measured in units of $t=1$.  We choose the Fr\"ohlich form,
\begin{align} 
\lambda(r)= \frac{ \lambda_0 \, e^{-r/\xi} } { (1+r^{2})^{3/2}}\,\,,
\label{eq:F}
\end{align}
which modulates the long-range interaction between electrons and phonons
with a screening length $\xi$ (measured in units of the lattice constant
$a=1$). In the $\xi \rightarrow 0$ limit, the local electron-phonon
coupling of the Holstein model is recovered.  In the $\xi \rightarrow
\infty$ limit, the interaction is reduced to a lattice version of
Fr\"{o}hlich model.  Throughout this paper we will consider the
half-filled situation where $\rho = (N_{\uparrow} + N_{\downarrow} )/N =
1$ with $N_\sigma$ the numbers of electrons of spin $\sigma$.  Although
we are generalizing the range of the electron-phonon interaction from
the $\xi=0$ Holstein limit, we will continue to consider only local
phonon degrees of freedom, i.e.~neglecting intersite interactions
between the phonons which would give rise to dispersive phonon
modes\cite{costa18}.

The (`pure') Hubbard  Hamiltonian (HH) $\hat {\cal K} + \hat {\cal P}_u$
is exactly soluble in 1D via the Bethe {\it ansatz}\cite{lieb68}.  For
all non-zero values of $U$, the ground state has spin order, with power
law decaying spin-spin correlations.  The Bethe {\it ansatz} solution
already interjects a cautionary note for numerical work: the gap which
immediately opens for $U>0$ is exponentially small.

The physics of the ground state of the extended Hubbard  
Hamiltonian (EHH)
$\hat {\cal K} + \hat {\cal P}_u + \hat{\cal P}_v$
has already been noted above:  CDW with true long
range order (owing to the discrete symmetry of the order parameter)
dominates for $U<2V$, with weaker, power law decaying SDW correlations at
$U>2V$ when the symmetry is continuous.  
The BOW phase consists of a pattern
in which the hopping 
$  \, \hat c^{\dagger}_{l, \sigma} 
\hat c^{\phantom{\dagger}}_{l+1, \,  \sigma} + 
 \, \hat c^{\dagger}_{l+1, \, \sigma} 
\hat c^{\phantom{\dagger}}_{l, \sigma} $ 
serves as a staggered order parameter,
alternating in amplitude for $l$ odd and even.
As for the CDW phase, 
the discrete nature of the breaking of the translational symmetry
in the BOW gives true long range order\cite{sengupta02}.

The Holstein Hamiltonian 
$\hat {\cal K} + \hat {\cal H}_{ph} + \hat {\cal P}_{ep}$
neglects the electron-electron interaction terms,
and also sets $\xi=0$ so that only the on-site
piece of $\hat {\cal P}_{ep}$ survives.
Similar to the HH, an (exponentially small) 
gap opens immediately for any $\lambda_0>0$ for
small phonon frequencies.  The existence of a gap for
large $\omega$ is still controversial, despite an impressively
large set of numerical and analytic investigations
\cite{hirsch83,caron84,schmeltzer87,bourbonnais89,wu95,takada96,hotta97,jeckelmann99,takada03,zhao10}.

The Holstein extended Hubbard (HEH) Hamiltonian
includes all the terms
$\hat {\cal K} + \hat {\cal H}_{ph} + \hat {\cal P}_u 
+ \hat{\cal P}_v + \hat{\cal P}_{ep}$,
again setting $\xi=0$.
The key added feature here is the possible existence of a gapless
metallic phase.  This occurs when the effective electron-electron
attraction $U_{\rm eff} = -\lambda_0^2 / \omega^2$ mediated
by integrating out the phonons, balances the on-site
repulsion $U$.
Contradictory results concerning
this gapless phase exist
\cite{clay05,hardikar07,takada03,takada96,fehske03,fehske04,fehske08,tezuka05,tezuka07,ejima10,chatterjee10}.

Finally, some work has been done on extensions of the HEH
which allow for non-local electron-phonon coupling.  As noted earlier,
Bonca and Trugman \cite{bonca01} and later Tam {\it et al.} \cite{tam14}
considered a situation in which a phonon mode exists on the bond 
between two sites, coupling to the sum of the fermionic density
on the two endpoints.
This work was based on the observation that, for the commensurate
densities under consideration here, diagonal long range orders, 
{\it i.e.} SDW and CDW, almost invariably dominate over
superconductivity.  The bond phonon mode will clearly favor the
formation of pairs on adjacent sites, opening the door to the possibility
of off-diagonal long range order.  Indeed, singlet superconductivity was
shown to occur in the small $U,V$ portion of the phase diagram.

Motivated by the observation that the long-range electron-phonon
interaction supports light polaron and bipolaron physics observed in
the cuprates, Hohenadler {\it et al.}\cite{hohenadler12} simulated the
fermion-boson model with the Fr\"{o}hlich interaction.  They found
that the extended interactions suppress the Peierls instability and
balance CDW and s-wave pairing.  With similar motivation, for
fullerenes and manganites, Spencer {\it et al.}\cite{spencer05}
studied the ground state properties of the screened Fr\"{o}hlich
polaron in weak, strong and intermediate coupling regions, as well
various screening lengths.

Following this review, we conclude this introduction with the
background and motivation for investigation of the full HEH with the
Fr\"{o}hlich form of the electron-phonon coupling (Eq.~\ref{eq:H}).
In the dilute limit, continuous-time quantum Monte Carlo (CTQMC) has
been used to study the effect of varying the range $\xi$ on polaron
and bipolaron formation\cite{hague09}.  The interest in studying
general momentum-dependent coupling constants is driven by a number of
factors.  First, a momentum dependent $\lambda(q)$ is implicated in
several experimental situations of considerable current interest,
including the origin of the ten-fold increase in the SC $T_c$ of FeSe
monolayers\cite{wang16}, the disparity in the values of the
electron-phonon coupling in SrTiO$_3$ inferred from tunneling below
$T_c$ and angle-resolved photoemission (ARPES) in the normal
state\cite{swartz16}, and the ``extended phonon collapse" in
2H-NbSe$_2$ \cite{weber11}.  Second, there are qualitative issues to
be addressed, {\it e.g.}~how the range of the electron phonon
interaction $\xi$ affects the competition between metallic and
Peierls/CDW phases at half-filling.  Here recent CTQMC studies in one
dimension have shown that as $\xi$ increases from zero, the metallic
phase is stabilized and, for sufficiently large $\lambda$, phase
separation (PS) can also occur \cite{hohenadler12}.  Finally, it has been
argued that recent improvements in the energy resolution of resonant
inelastic x-ray scattering (RIXS) have opened the possibility of an
experimental determination of the electron-phonon coupling across the
full Brillouin zone \cite{devereaux16}.  This observation carries the
exciting implication that materials-specific forms for $\lambda(q)$,
can be incorporated into QMC simulations of appropriate model
Hamiltonians.

Unfortunately, one cannot immediately study the simple case of a
longer range electron-phonon interaction in isolation.  The reason is
that when a fermion distorts the lattice for $\xi$ finite, it does so
at a collection of sites in its vicinity.  This distortion attracts
additional fermions, further increasing the distortion.  The resulting
cascade leads to a very high tendency to phase separation.  The
Holstein Hamiltonian ($\xi=0$) neatly evades this collapse since the
Pauli principle caps the fermion count on a displaced site.  Likewise,
the Bonca and Trugman model of bond phonons restricts the
electron-phonon coupling to a pair of sites.  The solution to this
dilemma in the general case is clear- the inclusion of non-zero
electron-electron repulsion.  A recent study which incorporates
on-site $U$ but retains $V=0$\cite{hebert18} revealed continued phase
separation over much of the phase diagram, with SDW occurring only if
$\xi$ and $\lambda_0$ were rather small.  We show here that $V$ can
significantly stabilize the CDW and SDW phases against collapse of the
density.

\section{2.  Computational Methodology}

Our technical approach is a straightforward generalization of the 
world-line quantum Monte Carlo (WLQMC) method for lattice fermions of Hirsch {\it et al.}
\cite{hirsch82}.  In this method, a path integral is written for the
partition function by discretizing the inverse temperature $\beta$
into intervals of length $\Delta \tau = \beta/L$.  
\begin{align}
{\cal Z} &= {\rm Tr} \big[ \, e^{-\beta \hat {\cal H} } \, \big]
\nonumber \\ &=
{\rm Tr} \big[ \, e^{-\Delta \tau \hat {\cal H}_1 } 
 \, e^{-\Delta \tau \hat {\cal H}_2 } 
\cdots
 \, e^{-\Delta \tau \hat {\cal H}_1 } 
 \, e^{-\Delta \tau \hat {\cal H}_2 } 
\, \big]
\label{eq:PI}
\end{align}
On the odd imaginary time
intervals, half of the Hamiltonian of Eq.~\ref{eq:H} acts,
\begin{align}
\hat {\cal H}_1 &=
\hat {\cal K}_1 + 
\frac{1}{2} \big( \,
\hat {\cal H}_{ph}  + 
\hat {\cal P}_u +
\hat {\cal P}_v +
\hat {\cal P}_{ep}
\, \big)
\nonumber \\
\hat {\cal K}_1  &=
-t \sum_{l\in {\rm odd}, \,  \sigma} 
\big( \, \hat c^{\dagger}_{l, \sigma} 
\hat c^{\phantom{\dagger}}_{l+1, \,  \sigma} + 
 \hat c^{\dagger}_{l+1, \, \sigma} 
\hat c^{\phantom{\dagger}}_{l, \sigma}  \big)
\label{eq:CB1}
\end{align}
while on the even intervals the other half acts,
\begin{align}
\hat {\cal H}_2 &=
\hat {\cal K}_2 + 
\frac{1}{2} \big( \,
\hat {\cal H}_{ph}  + 
\hat {\cal P}_u +
\hat {\cal P}_v +
\hat {\cal P}_{ep}
\, \big)
\nonumber \\
\hat {\cal K}_2  &=
-t \sum_{l \in {\rm even}, \, \sigma} 
\big( \, \hat c^{\dagger}_{l, \sigma} 
\hat c^{\phantom{\dagger}}_{l+1, \,  \sigma} + 
 \hat c^{\dagger}_{l+1, \, \sigma} 
\hat c^{\phantom{\dagger}}_{l, \sigma}  \big)
\label{eq:CB2}
\end{align}
The division of $\hat {\cal H}$ into two pieces necessitates the
presence of $2L$ time slices.

Complete sets of occupation number and phonon coordinate states 
are inserted between
each incremental time evolution operator.
The matrix elements are then evaluated, replacing 
all operators by space and imaginary time dependent c-numbers
corresponding to the eigenvalues of the intermediate states.
The fermion occupation number states and phonon coordinate
states are sampled stochastically 
by introducing local changes and accepting/rejecting
according to the ratio of matrix elements.

The utility of the ``checkerboard decomposition"\cite{hirsch82}
is that on each imaginary time slice the matrix element of the
incremental time evolution operator factorizes into a product of
independent two site problems.
The matrix elements are thus simple to evaluate.
The Monte Carlo moves are chosen to preserve local conservation laws
of the particle count on each two site plaquette.

The strengths of the WLQMC approach include a linear scaling in 
spatial system size $N$ and imaginary time $\beta$ (in contrast
to the $N^3$ scaling of auxiliary field QMC).
This is a consequence of the locality of the values of the
matrix elements.
More significantly, there is no fermion sign problem\cite{loh90,troyer05} 
at any filling 
as long as the hopping occurs exclusively 
between sites which are adjacent in the list of occupation numbers
labeling the state.

In this work we will be interested in the ground state properties
of Eq.~\ref{eq:H}, which we access by choosing $\beta$ sufficiently
large.  Our typical choice is $\beta \sim  N$.  We have checked that
we have reached the low $T$ limit, with the caveat noted earlier
that the exponential scaling of the gap 
precludes this at weak coupling.
In what follows,
we note when that concern affects our conclusions significantly.

The realization Eqs.~\ref{eq:PI}-\ref{eq:CB2}
of WLQMC works in the canonical ensemble, 
although it is straightforward to connect to grand-canonical
ensemble results by doing simulations on adjacent particle number
sectors and extracting the chemical potential
from a finite difference of ground state energies of
systems of different particle number\cite{batrouni90}.

There are some weaknesses to the method.  Most notably, it is
challenging to evaluate expectation values of operators which `break'
the world lines.  Thus access to single and two particle Green
function is restricted, unlike other methods like the
`worm'\cite{prokofev98} and stochastic Green function
(SGF)\cite{rousseau08} approaches.  In addition, WLQMC often suffers
from long autocorrelation times which are associated with the local
nature of the moves, and the tendency for world lines to have
significantly extended spatial patterns.  Both of these are addressed
in the worm and SGF methods, at, however, the cost of somewhat greater
algorithmic complexity.  Here, in $d=1$ due to the speed of the
approach, one can tolerate long autocorrelations simply by doing many
updates.

\begin{figure}[t]
\includegraphics[scale=0.33]{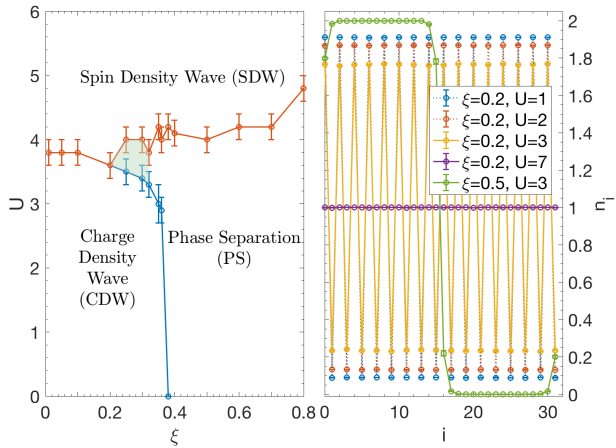}
\caption{(Color online) (a) Phase diagram of the HEH in the $U$-$\xi$
  plane at fixed $V=0$ and $\omega=0.5, \lambda_0=1.0$.  The lattice
  size is $N=32$ and inverse temperature $\beta=26$.  The procedure
  for determining the positions of the phase boundaries is
  described in subsequent figures and discussion, and in the right-hand
panel.
(b) Snapshots of 
    electron density $(n_{i} = n_{i, \uparrow}+n_{i, \downarrow})$
    profiles.  
For $\xi=0.5, U=3$, the particles are clumped in one region of
the lattice- the system exhibits PS.
For $\xi=0.2, U=7$, the total density is uniform, as expected for
an SDW where spin density oscillates, but charge density is constant. 
The three data sets at $\xi=0.2$ with $U=1,2,3$ are in the CDW phase.
The order parameter (size of charge oscillation) decreases as $U$ grows and
the phase boundary to the SDW is approached.
}
\label{fig:V00PD}
\end{figure}

In order to determine the phase diagram of Eq.~\ref{eq:H},
we monitor first a set of local observables: the fermion
kinetic energy, phonon kinetic energy, and fermion
double occupancy,
\begin{align}
{\cal K}_{el} &= -t \, \big\langle
\big( \, \hat c^{\dagger}_{l, \sigma} 
\hat c^{\phantom{\dagger}}_{l+1, \,  \sigma} + 
 \, \hat c^{\dagger}_{l+1, \, \sigma} 
\hat c^{\phantom{\dagger}}_{l, \sigma}  \big) \big\rangle
\nonumber \\
{\cal K}_{ph} &= \frac{1}{2} \,
\big\langle \, \hat P_{l}^2  \,\big\rangle
\nonumber \\
{\cal D} &=
\big\langle
\hat n_{l, \uparrow} \, \hat n_{l, \downarrow} 
\big\rangle
\label{eq:obs1}
\end{align}
In the absence of a broken CDW or BOW symmetry,
these are independent of lattice site $l$ since $\cal H$
is translation invariant.
Second, we evaluate the CDW and SDW structure factors,
\begin{align}
S_{\rm cdw} &= \frac{1}{N} \sum_{l,j}
\big\langle
\big( \, \hat n_{l, \uparrow} + \hat n_{l, \downarrow}  \, \big)
\, \big( \, \hat n_{j, \uparrow} + \hat n_{j, \downarrow}  \, \big)
\big\rangle
(-1)^{j+l}
\nonumber \\
S_{\rm sdw} &= \frac{1}{N} \sum_{l,j}
\big\langle
\big( \, \hat n_{l, \uparrow} - \hat n_{l, \downarrow}  \, \big)
\, \big( \, \hat n_{j, \uparrow} - \hat n_{j, \downarrow}  \, \big)
\big\rangle
(-1)^{j+l}
\nonumber \\
\label{eq:obs2}
\end{align}
to get insight into long range order (LRO).
These structure factors are independent of $N$ in a disordered
phase, but grow with $N$ in an ordered or
quasi-ordered phase, as the associated real space correlation
functions remain non-zero at large separations $|j-l|$.
Although the quantities of
Eq.~\ref{eq:obs1} are local and hence in principle not appropriate
to determining the appearance of non-analyticities associated
with transitions, we shall see
that they nevertheless show sharp signatures at the phase boundaries.

We also investigate the non-local observables including 
the charge and spin susceptibilities which involve
an additional integration over imaginary time,
\begin{align}
\chi_{\rm charge} &= \frac{\Delta\tau}{N} \sum_{\tau,l,j}
\big\langle
 \, \hat n_{l}(\tau)
 \, \hat n_{l}(0)
\big\rangle
(-1)^{j+l}
\nonumber \\
\hat n_l(\tau) 
&= \big( \, \hat n_{l, \uparrow} + \hat n_{l, \downarrow}  \, \big)(\tau) 
\nonumber \\
&= e^{\tau \hat {\cal H}} \,\,
\big( \, \hat n_{l, \uparrow} + \hat n_{l, \downarrow}  \, \big)(0)
\,\, e^{-\tau \hat {\cal H}} 
\nonumber \\
\chi_{\rm spin} &= \frac{\Delta\tau}{N} \sum_{\tau,l,j}
\big\langle
 \, \hat m_{l}(\tau)
 \, \hat m_{l}(0)
\big\rangle
(-1)^{j+l}
\nonumber \\
\hat m_l(\tau) 
&= \big( \, \hat n_{l, \uparrow} - \hat n_{l, \downarrow}  \, \big)(\tau) 
\nonumber \\
&= e^{\tau \hat {\cal H}} \,\,
\big( \, \hat n_{l, \uparrow} - \hat n_{l, \downarrow}  \, \big)(0)
\,\, e^{-\tau \hat {\cal H}}
\nonumber \\
\label{eq:Chispin}
\end{align}
The susceptibility provides a clearer signal of the SDW transition,
which is more subtle than the CDW transition owing to the continuous
nature of the symmetry involved.

\begin{figure}[t]
\includegraphics[height=7.0cm, width=8.0cm]{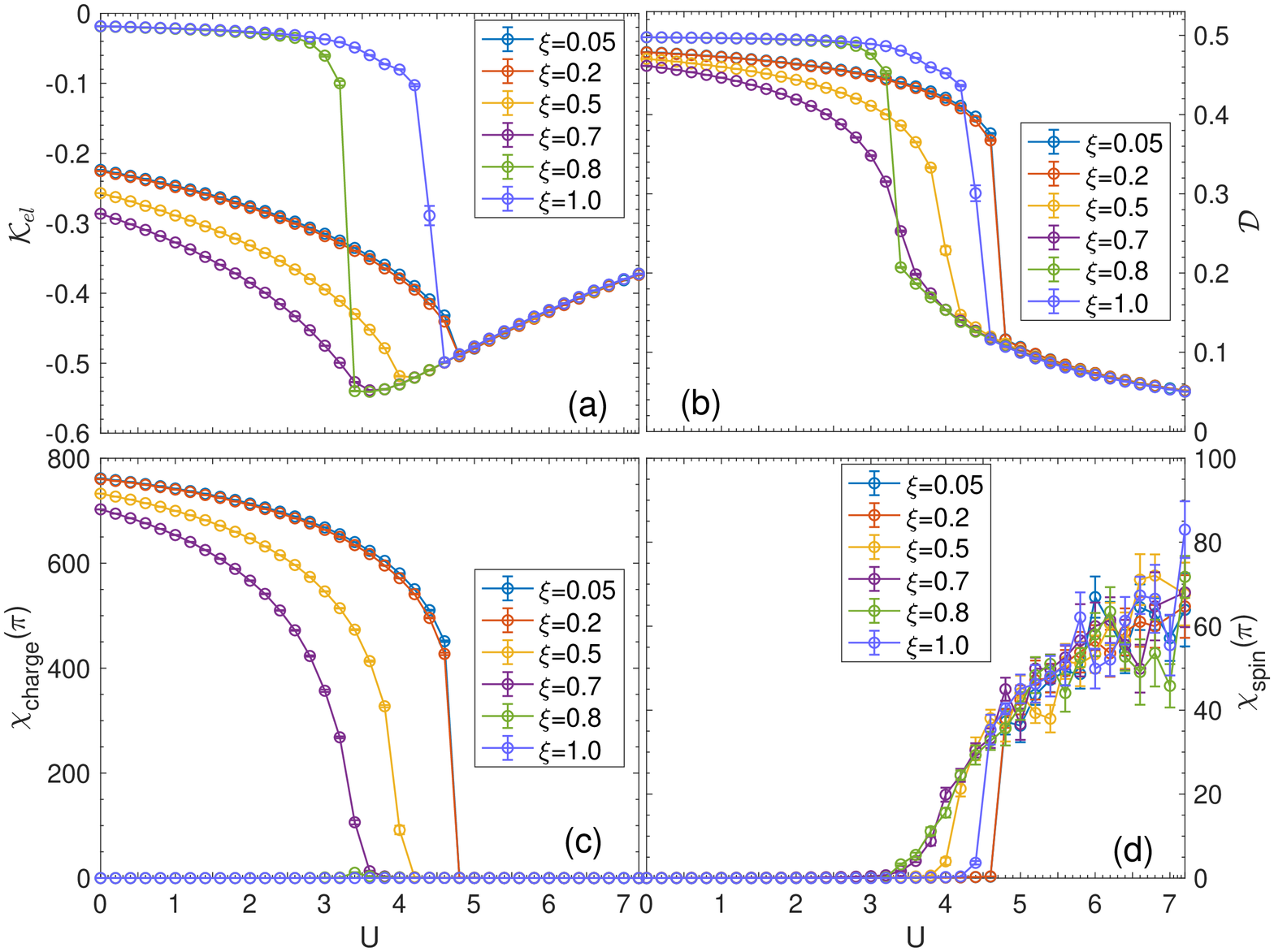}
\caption{(Color online) (a) Electron kinetic energy as a function of
  on-site repulsion $U$ for fixed intersite repulsion, $V=0.5$, and
  different values of electron-phonon interaction screening length
  $\xi$. (b) Double occupancy as a function of $U$.  A sharp drop in
  ${\cal D}$ and a sharp increase in the magnitude of ${\cal K}_{el}$
  remark the entry into the SDW phase from the large $\xi$ PS region
  as $U$ increases.  Structure at smaller $\xi$ is associated with the
  CDW-SDW boundary (See text and Fig.~\ref{fig:V05PD}(a)).  (c) Charge
  susceptibility as a function of $U$.  $\chi_{\rm charge}$ is big in the
  weak $U$ CDW phase, and drops at the SDW boundary.  (d) The spin
  susceptibility shows the opposite trend: $\chi_{\rm spin}$ is small at
  weak $U$, but increases as the SDW phase is entered at large $U$.
  The lattice size is $N=32$ and inverse temperature $\beta=26$.  The
  phonon frequency is fixed at $\omega=0.5$ and electron-phonon
  coupling strength $\lambda=1.0$.}
\label{fig:V05Support}
\end{figure}

\section{3.  Phase Diagrams at Fixed $V$} 

In this section we present the phase diagrams in the $U$-$\xi$ plane at
three fixed values of intersite electron-electron repulsion $V$,
focussing on the stabilization of the ordered phases
by $V$ against phase separation.  We begin by showing, in
Fig.~\ref{fig:V00PD}(a), the phase diagram for vanishing intersite
repulsion $V=0$ and with $\omega=0.5, \lambda_0=1.0$.  CDW
correlations, driven solely by the electron-phonon interaction (since
$V=0$), dominate at small $U$ and $\xi$, but are replaced by SDW as
$U$ increases, or by PS as $\xi$ increases.  Figure \ref{fig:V00PD}(a)
emphasizes the fragility of CDW order to PS: an electron-phonon
interaction range as short as a few tenths of a lattice spacing is
sufficient to drive the system into a regime where all electrons clump
together.  Although the errors bars make it somewhat uncertain, the
data in the range $0.2 \le \xi \le 0.35$ indicate the possibility of
penetration of a thin PS wedge at the SDW-CDW
boundary. 

Figure \ref{fig:V00PD}(b) provides direct
  visualization of total electron density on each lattice
  site. The alternating pattern in the CDW phase
  region reflects true long range charge order in the
ground state.  Quantum
  fluctuations due to the hopping $t$ 
reduce the magnitudes of the largest and the smallest
  densities from perfect double occupation ($n_i=2$)
and empty ($n_i=0$).
  For fixed screening length $\xi=0.2$, Fig. \ref{fig:V00PD}(b) shows
  the CDW phase is most stable at small $U$, with the charge oscillations
decreasing in size as $U$ grows, ultimately vanishing at $U=7$ in the
SDW.
  The figure also displays a subtle difference in 
the double occupancy ${\cal D}$ between the CDW and PS regions.  
Both have half their sites nearly empty and half nearly doubly occupied, 
but in the CDW case the $n_i$ values are farther from their
extreme limits $n_i=0,2$.  The reason is that the alternation
of empty and doubly occupied sites in the CDW allows for a much greater
degree of quantum fluctuations due to the hopping $t$ than
can occur in the PS state where the Pauli principle blocks
fermion mobility.

As noted in the introduction,
mitigating the strong tendency to PS at $V=0$ is what that motivates
our work here.  The results of Fig.~\ref{fig:V00PD} are consistent
with those obtained at $V=0$ in Ref.~\cite{hebert18}, which uses the
alternate stochastic Green function\cite{rousseau08} method.

We now turn to non-zero $V$. Our discussion will detail how
Fig.~\ref{fig:V00PD}, and subsequent phase diagrams, are obtained
through the analysis of the evolution of the observables of
Eqs.~\ref{eq:obs1}-\ref{eq:Chispin}.

\begin{figure}[t]
\begin{centering}
\includegraphics[scale=0.32]{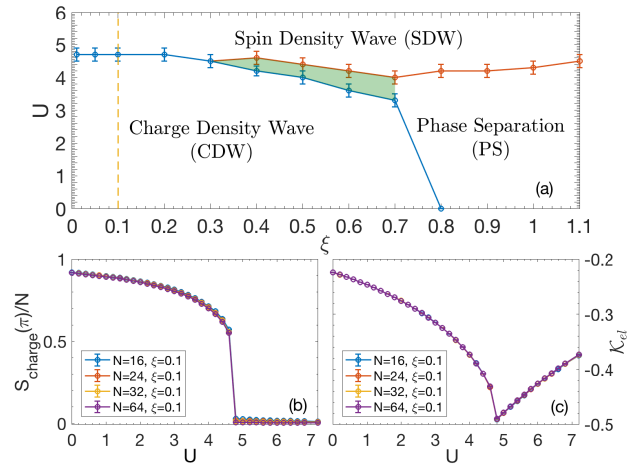}
\caption{(Color online) (a)Phase diagram of the HEH model in the $U$-$\xi$
  plane for $V=0.5$.  Compared to Fig.~\ref{fig:V00PD}(a) the CDW-PS
  line is pushed out from $\xi\approx 0.38$ at $V=0$ to $\xi\approx
  0.8$ here.  In the Holstein limit, CDW correlations are stabilized
  by an amount $2V$ against SDW.  The uncertainty in the precise
  nature of the phase where the CDW, SDW, and PS regions meet is
  indicated by the shaded region.  See text.  (b)Finite size effect 
  on the normalized charge structure factor 
  $S_{\rm charge}(\pi)/N$ at $\xi=0.1$. The magnitude of $S_{\rm charge}(\pi)$ 
  is proportional to the size of lattice, indicating the existence of 
  true long-range order in the CDW phase.  Therefore, the normalized 
  charge structure factor $S_{\rm charge}(\pi)/N$ is $N$ independent in the CDW phase.  
  $S_{\rm charge}(\pi)/N$ changes abruptly at 
  the same $U/t$ value for all $N$.  (c) Electron 
  kinetic energy $\mathcal{K}_{el}$ (per site) for different lattice sizes 
  at $\xi=0.1$.  The position of the cusp is N independent.}
\label{fig:V05PD}
\end{centering}
\end{figure}

Figure \ref{fig:V05Support}(a) shows the kinetic energy of the
electrons, ${\cal K}_{el}$, for a range of values of $\xi$ as a function
of the on-site $U$ for fixed $V=0.5, \omega=0.5,$ and $\lambda_0=1.0$.
For $\xi \gtrsim 0.8$, ${\cal K}_{el}$ is small until $U$ exceeds a
critical value.  This change is associated with a transition from a
small $U$ PS state, where the clumped electrons are unable to move
(except at the boundary of the occupied region) due to Pauli blocking,
to a large $U$ SDW state where alternating up and down occupation allows
considerable intersite hopping and hence large (in magnitude) ${\cal
K}_{el}$.  See Fig.~\ref{fig:V05PD}(a).  At strong $U$ we expect the
hopping to go as $t^2/U$ (second order perturbation induced by virtual
hopping), and indeed this fall-off fits the large $U$ data reasonably
well.  The value of the on-site repulsion $U$ which is needed to
eliminate PS is reduced as $\xi$ is lowered, as is expected since the
tendency for particles to clump is reduced as their interaction range is
shortened. 

In the small $\xi$ (Holstein) limit PS does not occur. Nevertheless
${\cal K}_{el}$ and ${\cal D}$ show appropriate signals of the CDW-SDW
transition: ${\cal  K}_{el}$ is largest in magnitude at the boundary
where the two insulators most closely compete, and ${\cal D}$ becomes
small as the SDW is entered.  The more rounded feature of ${\cal
K}_{el}$ at intermediate $\xi$ is one indication of the possible
intrusion of PS between CDW and SDW.  Again, see Fig.~\ref{fig:V05PD}(a).

The double occupancy ${\cal D}$ is given in Fig.\ref{fig:V05Support}(b).
To interpret it, we first note that ${\cal D}$ is not a good
discriminator between PS and CDW, because, in both, one has (subject to
quantum fluctuations) roughly half the sites doubly-occupied and half
empty, and hence ${\cal D} \sim 0.5$.  In the SDW phase, on the other
hand, ${\cal D} \sim 0$ since $U$ precludes double occupancy.  Thus the
most evident feature of Fig.\ref{fig:V05Support}(b) is the sharp drop in
${\cal D}$ as $U$ increases which occurs for all $\xi$.  These occur at
values consistent with the increase in the magnitude of ${\cal K}_{el}$
of Fig.\ref{fig:V05Support}(a).  ${\cal D}$ is not strictly zero in the
SDW phase because of the quantum (charge) fluctuations induced by $t$.
These gradually go down at $U$ increases.

Despite the fact that ${\cal D} \sim 0.5$ in both the CDW and PS states,
Fig.~\ref{fig:V05Support}(b) still shows a subtle signature of the
CDW-PS transition with increasing $\xi$ at fixed $U$: ${\cal D}$
initially falls as $\xi$ increases from $\xi=0.2$ to $\xi=0.7$, but then
rises again from $\xi=0.7$ to $\xi=1.0$.  The somewhat smaller values of
${\cal D}$ in the CDW result from fluctuations which are morely likely
when doubly occupied sites are surrounded by empty sites than in the PS
state where they are adjacent to each other.  This signal of the CDW-PS
boundary of Fig.~\ref{fig:V05PD}(a), although weak, is quite clear, and
lines up well with the small $U$ features in ${\cal K}_{el}$ in
Fig.~\ref{fig:V05Support}(a).

The spin susceptibility in Fig.~\ref{fig:V05Support}(d) reinforces the
inferences made from the local observables and confirms the large $U$
phase is SDW.  For each value of $\xi$, a sharp increase in
$\chi_{\rm spin}$ occurs as $U$ increases.  The critical value is not very
sensitive to $\xi$, changing from $U_c \sim 4.8$ at $\xi \sim 0.2$ to
$U_c \sim 4$ at $\xi \sim 0.8$.  This is reflected in the nearly
horizontal character of the phase boundary of Fig.~\ref{fig:V05PD}(a).
In the complete absence of any thermal or quantum fluctuations we have
$\chi_{\rm spin} = N\beta$ from Eq.~\ref{eq:Chispin}.  The values of
Fig.~\ref{fig:V05Support}(d) are an order of magnitude smaller:
Quantum fluctuation reduce correlations significantly, especially in
one dimension and for the case of a continuous order parameter which
has power law correlations at $T=0$.

Finally, the charge susceptibility is shown in
Fig.~\ref{fig:V05Support}(c).  
For $\xi \lesssim 0.7$, $\chi_{\rm charge}$ is large at small $U$, 
indicating the presence of CDW order
until PS and then SDW occur as $U$ increases.  For larger $\xi$ there
is no small $U$ CDW phase.  These observations are consistent with the
measurement of the local observables and spin susceptibility of
Figs.~\ref{fig:V05Support}(a), Figs.~\ref{fig:V05Support}(b) and
\ref{fig:V05Support}(d).

The full phase diagram at $V=0.5$ is given in Fig.~\ref{fig:V05PD}(a).
The PS regime has been pushed out to $\xi \sim 0.8$, in contrast to
$\xi \sim 0.38$ in Fig.~\ref{fig:V00PD}(a).  The SDW phase boundary is
remarkably flat: $U_c \sim 4-5$ regardless of the nature of the phase
(CDW or PS) beneath it.  For small $\xi$ the CDW region is also
increased in size vertically upward in going from $V=0$ to $V=0.5$.
This upward shift has been discussed by Hirsch\cite{hirsch83} in the
Holstein limit of local electron-phonon interaction: Nonzero $V$
combines with the tendency to form CDW order driven by the
electron-phonon interaction, producing a $2V$ change in the position
of the CDW-SDW boundary.  This estimate is in good quantitative
agreement with what we find in comparing Figs.~\ref{fig:V00PD}(a) and
\ref{fig:V05PD}(a).  In Fig.~\ref{fig:V05PD}(b) 
and Fig.~\ref{fig:V05PD}(c), we show the finite size effects on 
the normalized charge structure factor $S_{\rm charge}(\pi)/N$ 
and $\mathcal {K}_{el}$ at $\xi=0.1$.  
We have verified that the finite size effects for 
this parameter set are typical of those throughout phase space, 
allowing us to locate the phase boundaries accurately.

Just as the precise nature of the boundary between CDW and SDW in the
extended Hubbard model was challenging to uncover, with the original
picture of a direct transition (with a change from first to second
order at at tricritical point) being replaced by an intervening
BOW\cite{sengupta02}, in our studies the nature of a narrow region
where the CDW-SDW boundary meets PS is ambiguous.  The technical issue
is the difficulty in locating the precise positions of the CDW and SDW
transitions given the rounding effects of finite size lattices.  We
have indicated this uncertainty, which is greater as we turn on $V$,
by a shaded region in Fig.~\ref{fig:V05PD}(a).

\begin{figure}[t]
\begin{centering}
\includegraphics[scale=0.33]{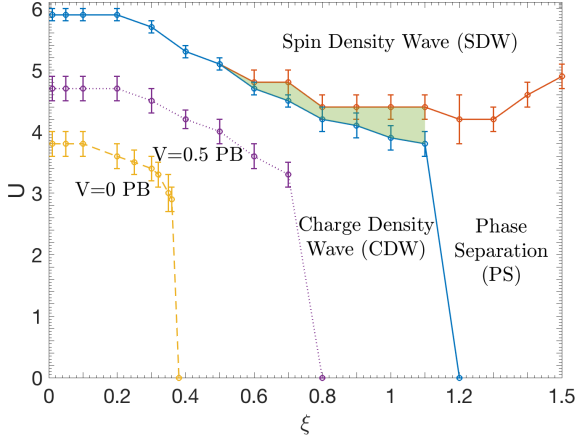}
\caption{(Color online) Phase diagram of the HEH model in the $U$-$\xi$
  plane for $V=1.0$ (solid lines).  The dashed and dotted lines
  indicate the phase boundary (PB) of the CDW region
  $V=0$ and $V=0.5$, with the SDW boundaries of
  Figs.~\ref{fig:V00PD}(a) and \ref{fig:V05PD}(a)
  suppressed for clarity.  As the intersite $V$ increases, the charge
  density wave is stabilized to longer range electron-phonon
  interaction range $\xi$.  At $\xi=0$ the CDW-SDW phase boundary
  moves upward approximately by $2V$ \cite{hirsch83}.  As with the
  preceding figure, the uncertainty in the precise nature of the phase
  where the CDW, SDW, and PS regions meet is indicated by the shaded
 region.  }
\label{fig:V10PD}
\end{centering}
\end{figure}

We will not show the detailed evolution of observables for larger
$V=1$, since their basic structure is the same as for $V=0.5$, just
discussed.  Figure \ref{fig:V10PD} is the resulting phase diagram.  PS
has been pushed to the region $\xi \gtrsim 1.2$. For smaller $\xi$,
while PS is absent, the increasing range of the el-ph coupling tilts
the CDW-SDW in favor of spin order.  In the Holstein limit, $\xi=0$,
the electron-phonon and nearest neighbor
electron-electron interactions work in concert to promote charge
order.  It is clear that as $\xi$ increases the electron-phonon
interaction no longer favors double occupation on the same site over
occupation of adjacent sites, leading to the downward slope of the CDW
boundary.

One might expect that as $\xi \rightarrow \infty$, the electron-phonon
interaction becomes independent of the fermion positions, and hence
that the CDW-SDW boundary should approach the usual $U=2V$ position.
The CDW boundary is indeed decreasing as $\xi$ grows, but for $V=1$
phase separation still occurs before the $V=2U$ values.  For large
$\xi$, the fact that the boundary of phase transition from CDW or PS
state to SDW state decreases compared to the Holstein limit suggests
that the effective attraction, $U_{eff}$, is smaller in Fr\"{o}hlich
model than in Holstein model.

\section{4.  Phase Diagrams at Fixed $U$} 

We now show a set of complementary phase diagrams in the $V-\xi$ plane
at fixed $U$.  Since their derivation lies in the same detailed
analysis of $\chi_{\rm charge}$, $\chi_{\rm spin}$, as well as ${\cal K}_{el}$
and ${\cal D}$, to that of the preceding section, we focus mostly on
the final phase diagrams.

\begin{figure}[t]
\hskip-0.20in
\includegraphics[scale=0.35]{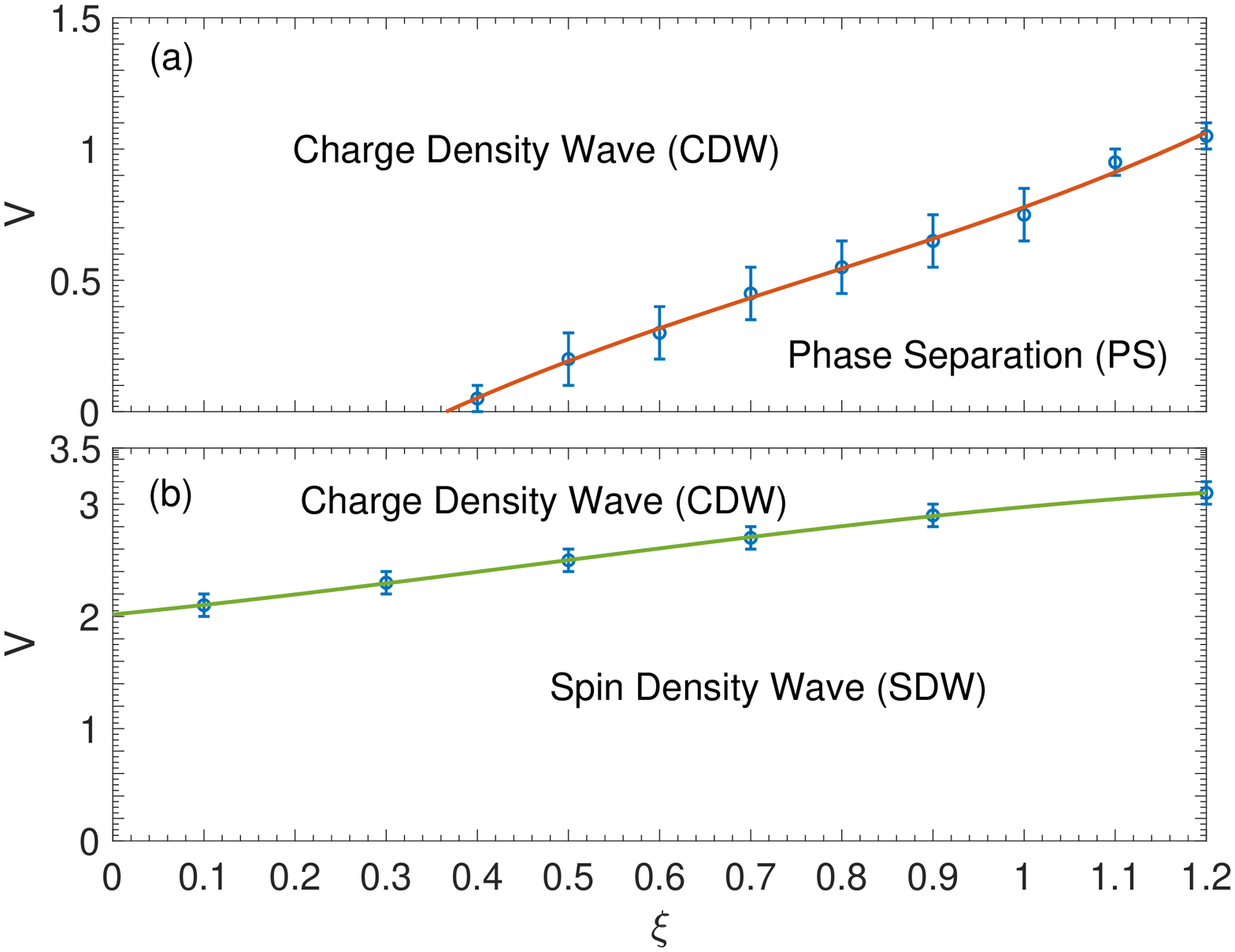}
\caption{(Color online) (a) Phase diagram of the HEH model in the $V$
  - $\xi$ plane for $U=3.0$.  The on-site repulsion is too weak to
  produce SDW.  (b) Phase diagram of the HEH model in the $V$ - $\xi$
  plane for $U=8.0$.  Here the on-site repulsion results in robust SDW
  phase and the combination of $U$ and $V$ eliminates all phase
  separation.  }
\label{fig:U3and8PD}
\end{figure}

We consider first small ($U=3$) and large ($U=8$) on-site repulsion,
which have the phase diagrams shown in Figs.~\ref{fig:U3and8PD}(a,b).
For $U=3$ the on-site repulsion is small enough that no SDW region
appears.  The intersite $V$ and electron-phonon interaction range
compete to give either CDW or PS.  For $U=8$, PS is replaced by SDW.
We can estimate the transition point in the Holstein limit:
$V=\frac{1}{2} (U - U_{eff}) = \frac{1}{2}(U - \lambda^2/\omega^2) =
2.0$, which agrees well with the $U=8$ phase diagram at $\xi=0$.
Despite its simplicity, Fig.~\ref{fig:U3and8PD}(b) carries a central
message of this paper: reasonable choices of $V$ and $U$ suppress the
PS which was noted in the first attempts to simulate momentum
dependent electron-phonon couplings.

\begin{figure}[t]
\includegraphics[height=7.0cm, width=9.0cm]{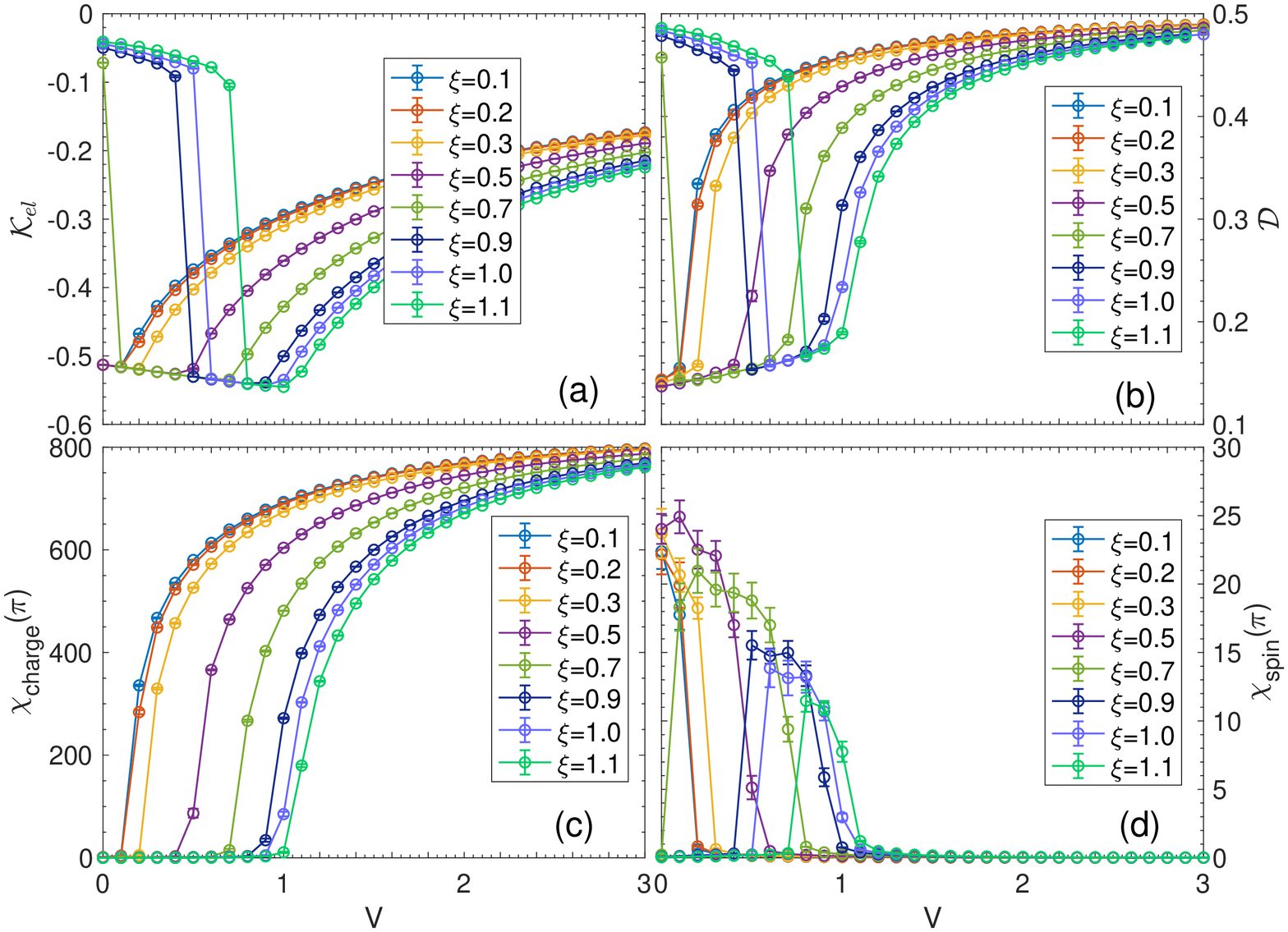}
\caption{(Color online) For the HEH model at $U=4$ and several values
  of $\xi$ we show as functions of $V$ the (a) electron kinetic
  energy, (b) the double occupancy, (c) the charge susceptibility, (d)
  the spin susceptibility. The number of site $N=32$ and inverse
  temperature $\beta=26$.  Phonon frequency $\omega=0.5$ and
  electron-phonon coupling strength $\lambda_0=1.0$.}
\label{fig:U3Support}
\end{figure}

The behavior at $U=4$ is considerably more complex and hence worth
discussing in more detail. Figure \ref{fig:U3Support}(a) indicates a
small electron kinetic energy until $V$ exceeds a threshold
value. Careful examination of the line shapes for $0.8 \lesssim \xi
\lesssim 1.1$ reveals that there are two regimes of large ${\cal
  K}_{el}$.  As $V$ is increased initially, ${\cal K}_{el}$ increases
abruptly in magnitude, and then continues a much slower increase.  At
a second critical $V$ a kink appears, there is a change in the sign of
the slope, and the magnitude of ${\cal K}_{el}$ begins to fall.  In
combination with the behavior of other observables, see below, we
interpret this to indicate a PS to SDW to CDW evolution with
increasing $V$ for these values of $\xi$.  At larger $\xi$ there is an
abrupt jump in ${\cal K}_{el}$, but then immediately a decrease in
magnitude from the newly large values.  For $\xi \le 0.7$, the absence
of small ${\cal K}_{el}$ indicates that there is no phase separation
region for these values of $\xi$.  These correspond to single PS to
CDW, and SDW to CDW transitions, respectively.

The the double occupancy curves (Fig.~\ref{fig:U3Support}(b)) look
rather similar to the electron kinetic energy.  Here, as discussed
earlier, low values of ${\cal D}$ are associated with SDW, while large
values can be either PS or CDW.  The existence of two separate
transitions, quite evident in ${\cal K}_{el}$ is less clear in ${\cal
  D}$.  However, close inspection of the $\xi=0.9, 1.0,$ and $1.1$
curves shows that ${\cal D}$ increases gradually with $V$ after
entering the SDW, but then abruptly changes slope for yet larger $V$.
This is consistent with changes from PS to SDW to CDW for these $\xi$.

Figures \ref{fig:U3Support}(c) and \ref{fig:U3Support}(d) give the
evolution of the charge and spin susceptibilities respectively.
Consistent with the data from the local observables, $\chi_{\rm spin}$
is big only in an intermediate range of $V$ for $\xi \gtrsim 0.7$,
below which lies PS and above which CDW.  For smaller $\xi$, SDW phase
extends all the way to $V=1.0$.  For all $\xi$, $\chi_{\rm charge}$
grows large above a critical $V$.  The shapes or the curves are not
markedly different for $\xi \lesssim 0.7$ where the CDW emerges from
SDW, and for $\xi \gtrsim 0.7$ where it emerges from a PS region.

\begin{figure}[t]
\includegraphics[scale=0.33]{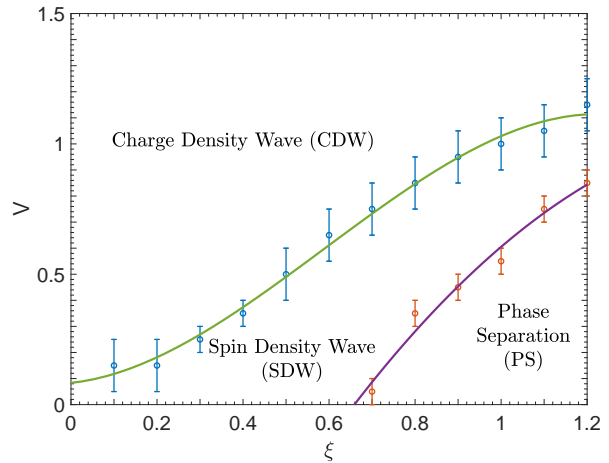}
\caption{(Color online) Phase diagram of the HEH model in the $V-\xi$ plane
for $U=4.0$ inferred from Fig.~\ref{fig:U3Support}.
The on-site repulsion is now big enough to give SDW 
sandwiched in between PS at large $\xi$ and CDW at
large $V$.  
}  
\label{fig:U4PD}
\end{figure}

Putting these plots together, we infer the $U=4$ phase diagram of
Fig.~\ref{fig:U4PD} which has CDW physics dominant at large $V$,
but two possibilities, SDW and PS, for the nature of the ground 
state at small $V$.  As with the constant $V=0.5$ phase diagram
of Fig.~\ref{fig:V05PD}(a) one thus encounters all three ground states as
one tunes the value of the electron-electron interactions.

\section{5.  Conclusions}

Recent attempts to include momentum dependent (finite range)
electron-phonon coupling without electron-electron interactions have
concluded that the tendency to phase separation dominates the
physics. This is mitigated to some extent\cite{hebert18} by an on-site
repulsion $U$, nevertheless PS is already dominant beyond relatively
small $\xi \gtrsim 0.5$.  This puts significant restrictions on the
nature of $\lambda(q)$ which would exhibit spatially homogeneous
densities- most of the weight would have to be at momenta $q > 2
\pi/\xi$, i.e.~well outside the first Brillouin Zone.

Here we have introduced a nearest-neighbor Coulomb interaction $V$ to
the extended Holstein-Hubbard Hamiltonian and obtained the resulting
phase diagrams at half-filling.  We have shown even relatively small
values of $V \sim \lambda_0$ can stabilize SDW and CDW phases and
eliminate (long period) density inhomogeneity.  Figure
\ref{fig:U3and8PD}(b) is a key result: PS is totally suppressed and
the effects of $\xi$ can be discerned in a context of global
thermodynamic stability.

Important questions remain open.  First, we have not attempted to
explore possible metallic phases. These are known to be extremely
challenging to address with QMC even in much more simple models, since
the gaps can vanish exponentially at weak coupling.  Second, we have
not doped the system.  
This paper has exclusively focused on half-filling.  
CDW and SDW phases tend to be optimized
at commensurate filling, but in $d=1$ the continued
presence of Fermi surface nesting can give
rise to order at $k<\pi$.
We generally expect superconductivity to be aided by some degree of
doping; it is already known to exist in models where bond
phonons couple to densities on both neighboring
sites\cite{bonca01,tam14}.  Much of the existing
experimental data on the materials which motivate the study of the model, including 
some organic superconductors and density wave systems, 
are away from half-filling.  Because the sign problem 
does not occur at any filling for our model in one dimension, 
it will be interesting to apply our method to study doped systems.  
As discussed here at half-filling,
a longer-range electron-phonon interaction
will play an important role in the competition between phase
separation and superconductivity, and we expect this
to be the case in the doped system as well.  

\section{Acknowledgements}

BX and RTS were supported by Department of Energy grant 
DE-SC0014671.
FH and GGB were supported by the French government, 
through the UCAJEDI Investments in the Future 
project managed by the National Research
Agency (ANR) with the 
reference number ANR-15-IDEX-01 and by 
Beijing Computational Science Research Center.



\begin{thebibliography}{100}

\bibitem{berger95}
E. Berger, P. Val\'a$\check{s}$ek, and W. von der Linden,
Phys. Rev. B {\bf 52}, 4806 (1995).

\bibitem{assaad07}
F. F. Assaad and T. C. Lang
Phys. Rev. B {\bf 76}, 035116 (2007).

\bibitem{nowadnick12}
E.A.~Nowadnick, S.~Johnston, B.~Moritz, R.T.~Scalettar, and
T.P.~Devereaux,
Phys. Rev. Lett. {\bf 109}, 246404.

\bibitem{johnston13}
S. Johnston, E.A. Nowadnick, Y.F. Kung, B. Moritz,
R.T. Scalettar, and T.P. Devereaux,
Phys. Rev. B {\bf 87}, 235133 (2013).

\bibitem{yamazaki14}
S. Yamazaki, S. Hoshino, and Y. Kuramoto,
JPS Conf. Proc. {\bf 3}, 016021 (2014).

\bibitem{macridin12}
A. Macridin, B. Moritz, M. Jarrell, and T. Maier,
J. of Phys.: Cond. Mat. {\bf 24}, (2012).

\bibitem{mendl17}
C. B. Mendl, E. A. Nowadnick, E. W. Huang, S. Johnston, B. Moritz, and
T. P. Devereaux,
Phys. Rev. B {\bf 96}, 205141 (2017).

\bibitem{ghosh18}
A. Ghosh, S. Kar, and S. Yarlagadda1a,
Eur. Phys. J. B {\bf 91}, 205 (2018).

\bibitem{karakuzu17}
S. Karakuzu, L.F. Tocchio, S. Sorella, and F. Becca,
Phys. Rev. B {\bf 96}, 205145 (2017).

\bibitem{karakuzu18}
S. Karakuzu and S. Sorella,
arXiv:1808.07759.

\bibitem{hirsch82}
J. E. Hirsch, R. L. Sugar, D. J. Scalapino, and R. Blankenbecler,
Phys. Rev. B {\bf 26}, 5033 (1982).

\bibitem{sengupta02}
P. Sengupta, A.W. Sandvik, and D.K. Campbell,
Phys. Rev. B {\bf 65}, 155113 (2002).

\bibitem{nakamura99}
M. Nakamura, 
J. Phys. Soc. Jpn. {\bf 68}, 3213 (1999)

\bibitem{nakamura00}
M. Nakamura, 
Phys. Rev. B {\bf 61}, 16377 (2000).

\bibitem{sandvik04}
A.W. Sandvik, L. Balents, and D.K. Campbell,
Phys. Rev. Lett. {\bf 92}, 236401 (2004).

\bibitem{tsuchiizu02}
M. Tsuchiizu and A. Furusaki,
Phys. Rev. Lett. {\bf 88}, 056402 (2002).

\bibitem{tsuchiizu04}
M. Tsuchiizu and A. Furusaki,
Phys. Rev. B {\bf 69}, 035103 (2004).

\bibitem{zhang04}
Y. Z. Zhang,
Phys. Rev. Lett. {\bf 92}, 246404 (2004).

\bibitem{tam06}
K-M. Tam, S-W. Tsai, and D.K. Campbell,
Phys. Rev. Lett. {\bf 96}, 036408 (2006).

\bibitem{bourbonnais08}
C. Bourbonnais and D. J\'erome,
{\it Physics of Organic Superconductors and Conductors},
A.G. Lebed (ed), Springer, Berlin (2008).

\bibitem{berlinsky79}
A.J. Berlinsky,
Rep. Prog. Phys. {\bf 42}, 1243 (1979).

\bibitem{gruner94}
{\it Density Waves in Solids},
G. Gruner,
Taylor and Francis, boca Raton (1994).

\bibitem{clay05}
R.T. Clay and R.P. Hardikar,
Phys. Rev. Lett. {\bf 95}, 096401 (2005).

\bibitem{hardikar07}
R. P. Hardikar and R. T. Clay,
Phys. Rev. B {\bf 75}, 245103 (2007).

\bibitem{tam07}
Ka-Ming Tam, S.-W. Tsai, D. K. Campbell, and A. H. Castro Neto
Phys. Rev. B {\bf 75}, 161103(R) (2007).


\bibitem{bonca01}
J. Bon$\check{c}$a and S.A. Trugman,
Phys. Rev. B {\bf 64}, 094507 (2001).

\bibitem{tam14}
K.M. Tam, S.W. Tsai, and D.K. Campbell,
Phys. Rev. B {\bf 89}, 014513 (2014).

\bibitem{costa18}
N.C. Costa, T. Blommel, W.-T. Chiu, G.G.  Batrouni, and R.T. Scalettar,
Phys. Rev. Lett. {\bf 120}, 187003 (2018).

\bibitem{lieb68}
E.H.~Lieb and F.Y.~Wu,
Phys. Rev. Lett. {\bf 20}, 1445 (1968).

\bibitem{hirsch83}
J.E.~Hirsch and E.~Fradkin,
Phys. Rev. B {\bf 27}, 4302 (1983).

\bibitem{caron84}
L.G. Caron and C. Bourbonnais,
Phys. Rev. B {\bf 29}, 4230 (1984).

\bibitem{schmeltzer87}
D. Schmeltzer,
J. Phys. C: Solid State Phys. {\bf 20}, 3131 (1987).

\bibitem{bourbonnais89}
C. Bourbonnais and L.G. Caron,
J. Phys. (France) {\bf 50}, 2751 (1989).

\bibitem{wu95}
C.Q. Wu, Q.F. Huang, and X. Sun,
Phys. Rev. B {\bf 52}, R15683 (1995).

\bibitem{takada96}
Y. Takada, 
J. Phys. Soc. Japan {\bf 65}, 1544 (1996).

\bibitem{hotta97}
T. Hotta and Y. Takada, 
Physica B {\bf 230}, 1037 (1997).

\bibitem{jeckelmann99}
E. Jeckelmann, C. Zhang, and S.R. White,
Phys. Rev. B {\bf 60}, 7950 (1999).

\bibitem{takada03}
Y. Takada and A. Chatterjee, 
Phys. Rev. B {\bf 67}, 081102 (2003).

\bibitem{zhao10}
J. Zhao and K. Ueda,
J. Phys. Soc. Japan {\bf 79}, 074602 (2010).

\bibitem{fehske03}
E. Fehske, A.P. Kampf, M. Sekania, and G. Wellein,
Eur. Phys. J. B {\bf 31}, 11 (2003).

\bibitem{fehske04}
E. Fehske, G. Wellein, G. Hager, A. Weisse, and
A.R. Bishop, 
Phys. Rev. B {\bf 69}, 165115 (2004).

\bibitem{fehske08}
E. Fehske, G. Hager, and E. Jeckelmann,
Europhys. Lett. {\bf 84}, 57001 (2008).

\bibitem{tezuka05}
M. Tezuka, R. Arita, and H. Aoki,
Phys. Rev. Lett. {\bf 95}, 226401 (2005).

\bibitem{tezuka07}
M. Tezuka, R. Arita, and H. Aoki,
Phys. Rev. B {\bf 76}, 155114 (2007).

\bibitem{ejima10}
S. Ejima and H. Fehske,
J. Phys: Conf. Ser. {\bf 200}, 012031 (2010)

\bibitem{chatterjee10}
A. Chatterjee,
Adv. Cond. Matt. Phys. 2010, 350787 (2010).

\bibitem{alexandrov04}
A.S. Alexandrov and B. Ya. Yavidov,
Phys. Rev. B {\bf 69}, 073101 (2004).

\bibitem{spencer05}
P.E. Spencer, J.H. Samson, P.E. Kornilovitz,
and A.S. Alexandrov,
Phys. Rev. B {\bf 71}, 184310 (2005).

\bibitem{hohenadler12}
M. Hohenadler, F.F. Assaad, and H. Fehske,
Phys. Rev. Lett. {\bf 109}, 116407 (2012).


\bibitem{hague09}
J.P. Hague and P.E. Kornilovitch,
Phys. Rev. B {\bf 80}, 054301 (2009).

\bibitem{wang16}
Y. Wang, K. Nakatsukasa, L. Rademaker, T. Berlijn, and
S. Johnston,
arXiv:1602.00656.

\bibitem{swartz16}
A.G. Swartz, H. Inoue, T.A. Merz, Y. Hikita,
S. Raghu, T.P. Devereaux, S. Johnston, and H.Y. Hwang,
arXiv:1608.05621.

\bibitem{hebert18}
F. H\'ebert, B. Xiao, V.G. Rousseau, R.T. Scalettar and G.G. Batrouni,
Phys. Rev. B {\bf 99}, 075108 (2019).

\bibitem{weber11}
F. Weber, S. Rosenkranz, J.-P. Castellan, R. Osborn,
R. Hott, R. Heid, K.-P. Bohnen, T. Egami, A.H. Said, and D. Reznik,
Phys. Rev. Lett. {\bf 107}, 107403 (2011).

\bibitem{devereaux16}
T.P. Devereaux, A.M. Shvaika, K. Wu,  K. Wohlfeld,  C.J. Jia,
Y. Wang,  B. Moritz, L. Chaix, W.-S. Lee,  Z.-X. Shen, G. Ghiringhelli,
and L. Braicovich,
arXiv 1605.03129. 

\bibitem{loh90}
E.Y.~Loh, J.E.~Gubernatis, R.T.~Scalettar,
S.R.~White,
D.J.~Scalapino, and R.L.~Sugar, 
Phys.~Rev.~B {\bf 41}, 9301 (1990).

\bibitem{troyer05}
M. Troyer and U-J. Wiese,
Phys. Rev. Lett. {\bf 94}, 170201 (2005).

\bibitem{batrouni90}
G.G. Batrouni, R.T. Scalettar, and G.T. Zimanyi,
Phys. Rev. Lett. {\bf 65}, 1765 (1990).

\bibitem{prokofev98}
N.V. Prokof\'ev, B.V. Svistunov, and I.S. Tupitsyn,
Phys. Lett. A {\bf 238}, 253 (1998).

\bibitem{rousseau08}
V.G. Rousseau,
Phys. Rev. E {\bf 77}, 056705 (2008).




\end{thebibliography}
\end{document}